\documentclass[twocolumn, times, tighten, appendixfloats]{aastex631}

\usepackage{amsmath}
\usepackage{romannum}
\usepackage{natbib}
\usepackage{enumitem}
\usepackage{soul}
\usepackage{longtable}

\usepackage{graphicx}           
\usepackage{latexsym}           
\usepackage{bm}                 
\usepackage[english]{babel}
\usepackage{color}              

\begin{document}

\pagenumbering{arabic}

\title{An Edge-on Regular Disk Galaxy at $z=5.289$}

\author[0000-0001-7592-7714]{Haojing Yan}
\affiliation{Department of Physics and Astronomy, University of Missouri, Columbia, MO 65211, USA}

\author[0000-0001-7957-6202]{Bangzheng Sun}
\affiliation{Department of Physics and Astronomy, University of Missouri, Columbia, MO 65211, USA}

\author[0000-0003-4952-3008]{Chenxiaoji Ling}
\affiliation{National Astronomical Observatories, Chinese Academy of Science, Beijing 100101, China}

\begin{abstract}

  While rotation-supported gas disks are known to exist as early as at
$z\approx 7$, it is still a general belief that stellar disks form late in the 
Universe. This picture is now being challenged by the observations from the 
James Webb Space Telescope (JWST), which have revealed a large number of 
disk-like galaxies that could be at $z>3$, with some being candidates at $z>7$. 
As an early formation of stellar disks will greatly impact our theory of galaxy
formation and evolution, it is important to determine when such systems first
emerged. Here we present D-CEERS-RUBIES-z5289 at $z=5.289\pm0.001$, the second
confirmed stellar disk at $z>5$, discovered using the archival JWST NIRCam 
imaging and NIRSpec spectroscopic data. This galaxy has a highly regular 
edge-on disk morphology, extends to $\sim$6.2~kpc along its major axis, and 
has an effective radius of $\sim$1.3--1.4~kpc. Such a large stellar disk is yet
to be produced in numerical simulations. By analyzing its 10-band spectral 
energy distribution using four different tools, we find that it has a high
stellar mass of $10^{9.5-10.0}M_\odot$. Its age is in the range of 
330--510~Myr, and it has a mild star formation rate of
10--30~$M_\odot$~yr$^{-1}$. While the current spectroscopic data do not allow
the derivation of its rotation curve, the width of its H$\alpha$ line from the
partial slit coverage on one side of the disk reaches $\sim$345~km~s$^{-1}$,
which suggests that it could have a significant contribution from rotation.

\end{abstract}


\section{Introduction}

  In the hierarchical structure formation paradigm, it is generally believed
that disk galaxies should form late in the cosmic history because major mergers,
which happened frequently in early times, would destroy disks. In addition,
active galactic nuclei and/or stellar feedback would also limit the formation
of disks. Indeed, stellar disks are usually seen at $z \approx 2$ and below,
which is in line with this picture. On the other hand, significant rotation in
extended gas (i.e., gas disk) has been observed at $z\approx 4$--7 
\citep[e.g.,][]{Hodge2012, Smit2018, Rizzo2020, Rizzo2021, Neeleman2020, 
Fraternali2021, Lelli2021, Tsukui2021, Pope2023, RO2023, Rowland2024}, which 
challenge this view. Encouragingly, recent zoom-in numerical simulations
suggest that disks could form early and be sustainable. For example, \
citet{Kretschmer2022} identified a cold gas disk at $z\approx 3.5$ in their
high-resolution simulations. \citet{Kohandel2024} showed that gas disks could
exist at $z>4$ and last for $\gtrsim 200$~Myr. \citet{Tamfal2022} and 
\citet{vanDonkelaar2024} simulated a Milky Way-size galaxy and found that a 
stellar disk could be in place as early as $z\approx 7$--8. Presumably, the
formation of stellar disks should lag behind that of gas disks because, in 
contrast to gas, stars are collisionless and harder to ``cool'' to form a disk.
Therefore, it is important to determine from the observational side when the 
first stellar disks emerged in the Universe. 

   Over the past 2 yr, a large number of candidate stellar disks have been 
found by the James Webb Space Telescope (JWST) at $z>3$
\citep[e.g.,][]{Fudamoto2022, Ferreira2022, Ferreira2023, Nelson2023a, 
Jacobs2023, Robertson2023b, HC2024, Kuhn2024}, which suggest an early formation 
of stellar disks. Spectroscopic confirmation of redshifts, however, is still
very limited. Using NIRCam images, \citet{Wu2023} identified a submillimeter
galaxy (SMG) at $z=3.06$ as a grand-design spiral, which is the first confirmed
stellar disk at $z>3$. Its redshift was previously obtained at the Atacama
Large Millimeter/submillimeter Array (ALMA). \citet{Nelson2023b} reported the
discovery of ``Twister-z5'', a fast-rotating galaxy at $z=5.3$ identified by
the NIRCam imaging and grism spectroscopic data. 

   In this paper, we present D-CEERS-RUBIES-z5289 (hereafter ``Dz5289'' for
short), a galaxy at $z=5.289$ with a very regular edge-on disk morphology. It
is among a large sample of disk galaxies that we visually classified using the
archival NIRCam data (B. Sun et al., in prepparation), and its accurate
redshift was based on NIRSpec spectroscopy. The paper is organized as follows.
We describe its morphology and photometry in Section 2 and its spectroscopy in
Section 3. The analysis of its stellar population is given in Section 4. We
present a brief discussion in Section 5 and conclude with a summary in Section
6. All magnitudes are in the AB system. We adopt the following cosmological 
parameters throughout: 
$H_0=71$~km~s$^{-1}$~Mpc$^{-1}$, $\Omega_M=0.27$ and $\Omega_\Lambda=0.73$.

\section{Imaging Data}

   The most critical imaging data are those taken by the Cosmic Evolution Early 
Release Science \citep[CEERS, PID 1345;][]{Finkelstein2023} using 
NIRCam. The location of our target is covered in seven 
passbands: F115W, F150W, and F200W in the short wavelength (SW) channel, 
and F277W, F356W, F410M, and F444W in the long wavelength (LW) channel.
We reduced these data on our own using the JWST science calibration pipeline 
(hereafter referred to as the ``pipeline'') version 1.9.4 in the processing
``context'' of {\tt jwst\_1046.pmap}, and the details of the procedures were 
described in \citet[][]{Yan2023e}. For this work, we created the mosaics at the
pixel scale of 0\farcs06. We also made use of the Hubble Space Telescope (HST)
data from the Cosmic Assembly Near-Infrared Deep Extragalactic Legacy Survey
\citep[CANDELS;][]{Grogin2011, Koekemoer2011}, which
are the ACS F606W and F814W and the WFC3 F125W images. Our source is located at
R.A.~=~14:19:38.33 and decl.~=~52:55:23.10 (J2000.0). Table~\ref{tab:phot}
summarizes the effective exposure times at this location in all these bands.

\begin{figure*}
    \centering
    \includegraphics[width=\textwidth,height=\textheight,keepaspectratio]{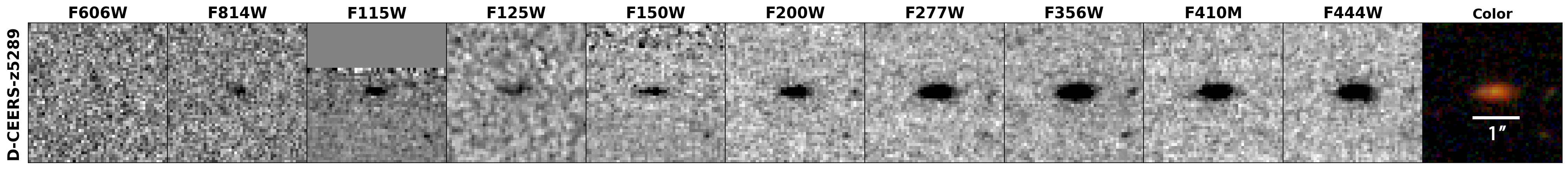}
    \caption{Image stamps of Dz5289 (3\arcsec$\times$3\arcsec\ in 
    size). The bands are as labeled. The F606W, F814W and F125W images are from 
    the HST CANDELS program, while all others are from the NIRCam observations 
    of the JWST CEERS program. The color image uses F606W + F814W + F115W + 
    F125W + F150W as blue, F200W + F277W as green, and F356W + F410M + F444W as 
    red. As the white line indicates, the galaxy extends at least 1\farcs0 along
    its long axis.
    }
    \label{fig:stamps}
\end{figure*}

\subsection{Morphology}

   Figure~\ref{fig:stamps} shows the stamp images of our target in the 
aforementioned JWST and HST passbands as well as the color composite using 
these data. Its edge-on disk morphology is clearly revealed in the four 
NIRCam LW bands, and the disk extends to at least $\sim$1\farcs0 along its major
axis. To obtain quantitative assessments of its morphology, we utilized the 
{\sc Galfit} software \citep[][]{Peng2002, Peng2010} to model this galaxy in 
these four bands.
  
   Running {\sc Galfit} needs the point spread function (PSF) of the image to
be analyzed. We constructed the empirical PSFs 
following the procedures of \citet[][]{LY2022}. For a given band, we selected 
isolated stars and made cutouts of $201\times 201$ pixels centered on them. The
sources around the stars were masked, and the cutout images were subsampled to 
a finer grid by 10 times. The centers of these cutouts were also aligned in 
this process. The fluxes of the stars were normalized to unity, and the 
normalized images were stacked using median. The stacked star image 
was rebinned by a factor of 10 in both dimensions to restore the original 
resolution. Finally, we cut out the central $101 \times 101$ pixels of the 
rebinned image, which was adopted as the PSF image. While {\sc Galfit} would 
only be run on the LW images, the PSFs were generated for all the bands as they 
would be used for photometry (see Section 2.2).

    The {\sc Galfit} modeling was done on 6\arcsec~$\times$~6\arcsec\ cutouts,
which contain some unrelated sources and contaminants. These could affect the
modeling and must be removed. For each image, we estimated the mean value and 
rms of the background using \texttt{sigma\_clip} in {\sc Astropy} and then 
created a pseudo background map based on the mean and rms values. Any undesired 
pixels were then replaced by those from the pseudo background map. 
{\sc Galfit} was run on the final images that contain the target galaxy at the 
center with a clean background that preserves the original noise property.

\begin{figure*}
    \centering
    \includegraphics[width=0.95\textwidth,height=\textheight,keepaspectratio]{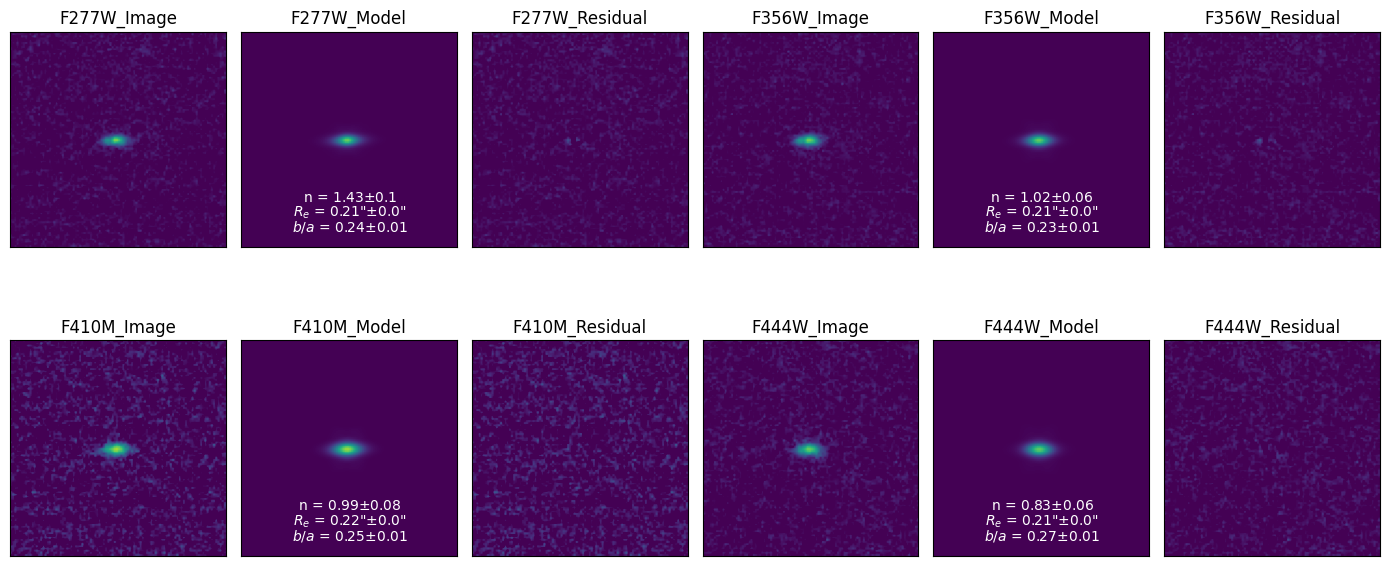}
    \caption{{\sc GALFIT} fitting results in NIRCam F277W, F356W, F410M, and F444W, using a single S\'ersic profile. For each band, the original image, the best-fit model from GALFIT and the residual image after the model subtraction are shown from left to right. The values of the S\'erscic index $n$, effective radius $R_e$, and axis ratio $b/a$ are labeled on the model image.
    }
    \label{fig:galfit}
\end{figure*}

   This galaxy was successfully fitted by using a single S\'ersic profile
\citep[][]{Sersic1963} in the four NIRCam LW bands. As shown in 
Figure~\ref{fig:galfit}, the results in these bands are consistent with the 
S\'ersic index $n$ of 0.83--1.43, the effective radius $R_e$ of 
0\farcs21--0\farcs22, and the axis ratio $b/a$ of 0.23--0.27.
These results support that our object is an edge-on disk galaxy.

\subsection{Photometry}

   Our NIRCam image products are registered to the astrometry of CANDELS, and
all the images used in this work are aligned in a pixel grid. These images were 
all PSF-matched to the angular resolution of the F444W image by convolving them 
with the convolution kernels created from the PSF images (see Section 2.1). We 
ran {\sc SExtractor} \citep[][]{Bertin1996} in the dual-image mode for 
photometry, using the F444W image for detection. We adopted the isophotal 
magnitudes (``MAG\_ISO'') and estimated the errors using the rms maps that we 
derived using the {\sc astroRMS} software tool
\footnote{Courtesy of M. Mechtley; see \url{https://github.com/mmechtley/astroRMS}}. The results are reported in Table~\ref{tab:phot}. For this isolated 
source of moderate brightness, the differences between MAG\_ISO magnitudes and
the total magnitudes (``MAG\_AUTO'') are negligible. Our object is not detected
in the HST ACS F606W band, for which we quote the 2$\sigma$ limit calculated on
the F606W rms map within the area of the same size as the isophotal aperture
used in other bands.

   We note that the NIRCam F115W data have to be discarded because the source
is too close to the field edge in this band (see Figure~\ref{fig:stamps}), 
which leads to large errors in photometry. Fortunately, the CANDELS HST WFC3 
F125W image samples the similar wavelength range and compensates the loss.

\begin{table}[hbt!]
    \centering
    \caption{Summary of imaging data and photometry.}
    \begin{tabular}{ccc} 
        Filter & Exposure (s) & AB mag. \\ \hline
        F606W & 5066 & $\geq28.72$ \\ 
        F814W & 11,742 & $27.29\pm0.26$ \\ 
        F115W & 1894 & ... \\
        F125W & 1813 & $26.49\pm0.18$ \\ 
        F150W & 1119 & $26.63\pm0.28$ \\ 
        F200W & 4751 & $26.17\pm0.05$ \\ 
        F277W & 3083 & $25.63\pm0.04$ \\ 
        F356W & 3082 & $25.52\pm0.04$ \\ 
        F410M & 3082 & $25.12\pm0.05$ \\ 
        F444W & 5650 & $25.33\pm0.02$ \\ 
        \hline 
    \end{tabular}
    \raggedright
    \tablecomments{
    The upper limit for F606W is the 2~$\sigma$ limit based on the rms map,
    calculated within a circle aperture equivalent to the isophotal area as 
    defined in F444W and centered at the source location. 
    The photometry in F115W is discarded because it is unreliable (due to the 
    object being too close to the field edge in this band).
    }
    \label{tab:phot}
\end{table}

\section{NIRSpec Spectroscopy}

   We used the public data from the RUBIES program (PID 4233, PIs: de Graaff 
\& Brammer; \citealt{deGraaff2024b}), which is a spectroscopic survey in 
multiple fields (including CEERS) utilizing the NIRSpec multiobject 
spectroscopy mode with the microshutter assembly (MSA). The observations were
done in a three-slitlet configuration of open shutters with a three-point
nodding pattern, and the data were taken in both the PRISM/CLEAR (hereafter 
PRISM) and G395M/F290LP (hereafter G395M) disperser/filter setups with the same
slit orientation. Under these setups, the spectral resolutions are $R\sim100$
and $\sim1000$, and the wavelength coverages are 0.6--5.3~$\mu$m and
2.87--5.10~$\mu$m, respectively. Overall, the total exposure time was 
$\sim$2.84~ks in both setups. 

   To reduce these data, we first retrieved the Level 1b products from MAST 
and processed them through the {\tt calwebb\_detector1} step of the pipeline 
(version 1.14.0) in the context of {\tt jwst\_1234.pmap}. The output 
``rate.fits'' files were then processed through the {\sc msaexp} package 
\citep[][version 0.8.4]{Brammer_msaexp2023}, which provides an end-to-end 
reduction including the final extraction of spectra. Briefly, the procedure
removes the so-called ``$1/f$'' noise pattern, detects and masks the 
``snowball'' defects, subtracts the bias level, applies the 
flat-field, corrects the path-loss, does the flux calibration, traces spectra 
on all single exposures, and combines the single exposures with outlier 
rejection. The background subtraction is done using the measurement in the 
nearest blank slit.

\begin{figure*}
    \centering
    \includegraphics[width=1.0\textwidth,keepaspectratio]{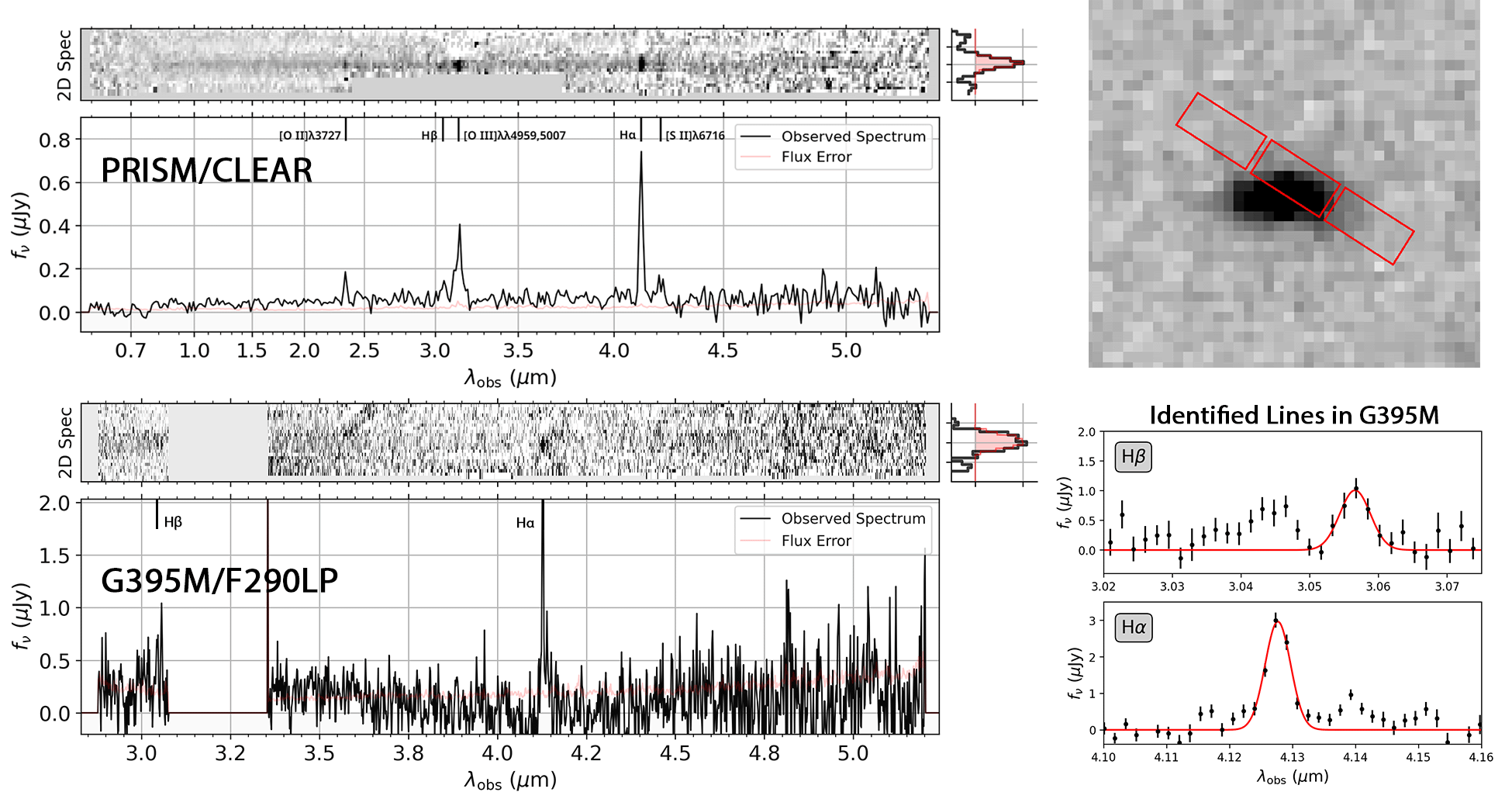}
    \caption{Results of the NIRSpec spectroscopy. The left two panels display 
    the 2D and 1D spectra from the PRISM and G395M observations, respectively.
    In each 1D-spectrum plot, the spectrum is shown in black and the error is
    shown in light red. The identified lines are labeled. The histogram
    attached to the right side of each 2D spectrum shows the cross-dispersion
    direction plot of the inverse-weighted sum of the 2D spectrum in the
    dispersion direction, and the 1D spectrum is generated by weight-averaging
    pixels along the cross-dispersion direction with weights as described by
    the histogram. The upper-right panel shows the slit (red rectangles)
    overlaid on the F444W image (2\farcs3$\times$2\farcs3 in size); only part
    of the galaxy was covered by the slit. The lower-right panel is a zoom-in
    view to the regions of the H$\alpha$ and H$\beta$} emission lines in the
    G395M 1D spectrum, where the red line represents the best-fit Gaussian
    profile to the data (black plus symbols). 
    \label{fig:spec}
\end{figure*}

   Our target is the source number 64132 in the RUBIES program.
As shown in Figure~\ref{fig:spec}, it was only 
partially covered by the slit. Nevertheless, the data are of sufficiently high
quality to allow for the identification.
The extracted 2D and 1D spectra are shown in Figure~\ref{fig:spec}.

   From the PRISM data, we identified five emission lines with a 
signal-to-noise ratio (S/N) $\approx 3.5$--18.5: 
[O~\textsc{\romannum{2}}]\,$\lambda$3727, H$\beta$, 
[O~\textsc{\romannum{3}}]\,$\lambda\lambda$4959,5007 doublet (unresolved), 
H$\alpha$, and [S~\textsc{\romannum{2}}]\,$\lambda$6716. Their observed 
wavelengths are 2.35, 3.06, 3.15, 4.13 and 4.22~$\mu$m, respectively, which put 
its redshift firmly at $z=5.29\pm0.01$. Interestingly, the Ly$\alpha$ emission
line is not visible, although the Lyman-break signature is clearly present. 
We fitted a 1D Gaussian profile to each of the five emission lines and measured
the line intensities within $2\times$FWHM from the central wavelength, and the
obtained values are listed in Table~\ref{tab:emlines}. A slight complication
was that only one exposure was available over the range of 2.4--3.7~$\mu$m: one
exposure had no data in this range due to some unknown reasons, and one other
had to be discarded because of the severe contamination. As a result, the
spectrum in this section is contributed by only a single exposure (i.e., with
only one-third of the total integration time), and some contaminants cannot be
removed. Upon close examination, we found that the H$\beta$ line was affected
by such a contaminant very nearby. To remedy this, we performed the Gaussian
fit of this line by setting the initial guess of its central wavelength based
on the position of the H$\alpha$ line and that of its FWHM to 0.001~$\mu$m. The 
H$\beta$ line was successfully separated from the contaminant in this way, 
although its intensity measurement inevitably had a large error.

   In the G395M data, the range of
$3.08\lesssim \lambda_{\rm obs}\lesssim 3.32\mu$m is not covered because it 
falls in the gap in between the detectors. In addition, we had to exclude one
exposure for the part bluer than 3.08~$\mu$m due to a possible contamination.
In the final spectrum, we were able to identify the strong H$\alpha$ in this
higher-resolution setting.  With the observed wavelength of 
4.127\textpm0.001~$\mu$m of this line, the redshift of the galaxy is refined to
5.289\textpm0.001. Using this precise redshift, we were also able to identify
the H$\beta$ line at 3.057~$\mu$m. The FWHM of the H$\alpha$ line is 
47.5\textpm2.9\AA, which translates to
$\Delta v_{\rm H\alpha}=345\pm21$~km~s$^{-1}$. 
This reflects the contributions from both the rotation and the random motion in
the part of the galaxy covered by the slit; however, these two components 
cannot be separated in the current data. The line intensity was measured in the
same way as in the PRISM data, and the value is also reported in 
Table~\ref{tab:emlines}. The measurements from both settings agree within the 
errors. The weaker [S~\Romannum{2}]$\lambda$6716 line was not detected in this
setup, which is likely due to its lower sensitivity as compared to the PRISM
setup.

\begin{table}[hbt!]
    \centering
    \caption{Line intensity measurements.}
    \begin{tabular}{cc} \hline
       Emission Line  & Intensity \\
                      & ($10^{-18}$~erg~s$^{-1}$~cm$^{-2}$) \\ \hline
        [O~\textsc{\romannum{2}}]\,$\lambda$3727 & 2.03\textpm0.35 \\ 
        H$\beta$ & 0.59\textpm0.16 \\
                  & 0.61\textpm0.15$^\dagger$ \\   
                  
        [O~\textsc{\romannum{3}}]\,$\lambda\lambda$4959,5007 & 3.78\textpm0.29 \\ 
        H$\alpha$ & 2.40\textpm0.13 \\ 
                  & 2.61\textpm0.18$^\dagger$ \\
                  
        [S~\textsc{\romannum{2}}]\,$\lambda$6716 & 0.52\textpm0.15 \\ 
        \hline
    \end{tabular}
    \raggedright
    \tablecomments{The values for H$\alpha$ and H$\beta$ marked with 
    $\dagger$ are obtained from the G395M observations, while all others are 
    based on the PRISM observations.}
    \label{tab:emlines}
\end{table}

    As we have the measurements for both H$\alpha$ and H$\beta$, we can derive
the extinction value through their line ratio, bearing in mind the caveat that
the result is subject to the large error of the H$\beta$ line. The observed
ratio between H$\alpha$ and H$\beta$ is $R_{\rm obs}=4.07\pm1.12$ from the
PRISM data (the ratio from the G395M data is similar), while the intrinsic
ratio in Case B recombination is $R_{\rm int}=2.86$. Using the extinction curve
of \citet[][]{Calzetti2000}, the reddening is 
\begin{align}
    {\rm E(B-V)_{gas}}=\frac{\log_{10}(R_{\rm obs}/R_{\rm int})}{0.4(k_{\rm H\beta}-k_{\rm H\alpha})}=0.30\pm0.20, 
\end{align}
where $k_{\rm H\alpha}=3.327$ and $k_{\rm H\beta}=4.598$. The extinction at the
location of the H$\alpha$ line is therefore 
$A({\rm H\alpha})=k_{\rm H\alpha}\times {\rm E(B-V)}=1.00\pm0.67$. After
correction for the extinction, the intrinsic intensity of the H$\alpha$ line 
is $(6.03\pm3.68)\times 10^{-18}$~erg~s$^{-1}$~cm$^{-2}$, which corresponds to 
the line luminosity of $L_{\rm H\alpha}=(1.86\pm1.14)\times10^{42}$~erg/s.
If using the G395M measurement of the line and applying the same correction, 
its intrinsic intensity is 
$(6.53\pm 4.01)\times 10^{-18}$~erg~s$^{-1}$~cm$^{-2}$ and the line luminosity 
is $(2.03\pm1.24)\times10^{42}$~erg/s. Note that the slit only covered part of 
the galaxy, and therefore these values are only lower limits; if the galaxy is
uniform in its H$\alpha$ line emission, the total values should be at least
twice as high.

   Using the formalism of \cite{Kennicutt1998a} that converts the H$\alpha$ line
luminosity to the instantaneous star formation rate (SFR), the values quoted
above correspond to SFR$_{\rm H\alpha}$ of 14.8$\pm$9.0 and 
16.1$\pm$9.8~$M_\odot$/yr, respectively. Again, these are only lower limits due
to the partial slit coverage.

\section{Spectral Energy Distribution Analysis}
    
    To further study the stellar populations of Dz5289, we fitted its spectral
energy distribution (SED) constructed from the photometry reported in
Table~\ref{tab:phot}. In order to test the robustness of the results, we
employed four different fitting tools: 
{\sc EAZY}-py \citep[][version 0.6.8]{Brammer2008},
{\sc CIGALE} \citep{Burgarella2005, Noll2009, Boquien2019} version 2022.1
\citep{YangCIGALE2022},
{\sc Bagpipes} \citep[][version 1.0.4]{Carnall2018}, and {\sc Prospector} 
\citep[][version 1.3.0]{Leja_prospector2017, Johnson_prospector2021}. 
The fitting was done at the fixed redshift of 5.29, and we used the Calzetti
extinction law.

    In the {\sc EAZY}-py run, we used the ``tweak\_fsps\_QSF\_12\_v3''
templates, which are a modified version of the flexible stellar population
synthesis models \citep[FSPS;][]{Conroy2010} tailored for galaxies at high
redshifts \citep{Finkelstein2022a, Larson2023}. These templates use the initial
mass function (IMF) of \citet{Kroupa2001}. 

    For {\sc CIGALE}, we used the stellar population synthesis models of 
\citet[][``BC03'']{Bruzual2003} with the IMF of 
\citet{Chabrier2003}. We adopted the delayed-$\tau$ star formation history 
(SFH), where SFR~$\propto te^{-t/\tau}$ and $\tau$ could vary from 0 to 10~Gyr.
The metallicity was freed within a set of numbers available in the program, with
$Z$ ranging from 0.0001 to 0.05. We also added contribution from nebula 
emission, and the ionization parameter was set to vary in 
$-4\leq\log(U)\leq-2$. The dust reddening parameter ${\rm E(B-V)_{\rm star}}$ 
could vary from 0 to 1.5~mag. 

    {\sc Bagpipes} utilizes the BC03 models with the Kroupa IMF. We ran it
using the exponentially decaying SFH (the ``$\tau$ model''), which is in the
form of SFR~$\propto e^{-t/\tau}$. We set $\tau$ as a free parameter varying
between 10~Myr and 15~Gyr. The metallicity was allowed to vary in the range of 
$0\leq Z/Z_\odot\leq 2.5$, and $A_{\rm V}$ could change from 0 to 8~mag.
We also included the nebula emission lines and allowed the ionization parameter 
to vary in the range of $-4.0\leq\log(U)\leq2.0$. 

    In the {\sc Prospector} run, we used the series of FSPS models of the
delayed-$\tau$ SFH available in its template library.
We adopted the same parameters as in the {\sc Bagpipes} run.

\begin{figure*}[hbt!]
    \centering
    \includegraphics[width=\textwidth]{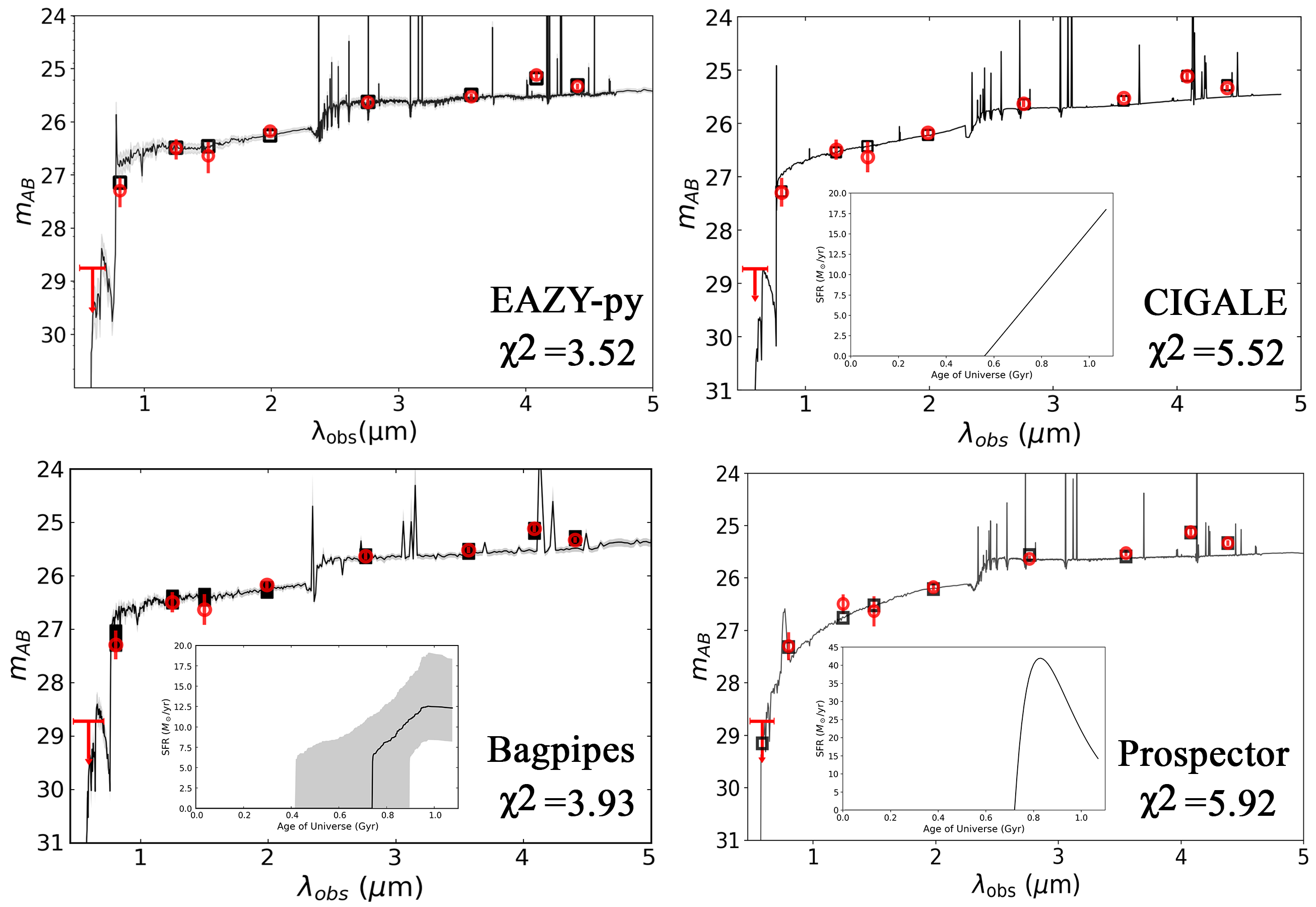}
    \caption{Fitted template spectrum (black curve) superposed on the SED (red
    symbols) in four fitting runs using different tools as labeled. The black
    squares are the synthesized magnitudes from these template spectra. The 
    corresponding $\chi^2$ values are also noted. For {\sc EAZY}-py, the
    spectrum is that of the best-fit template. For {\sc CIGALE, Bagpipes} and 
    {\sc Prospector}, it is the 50th percentile posterior spectrum. The 
    evolution of SFR (as a function of time; 50th percentile posterior value)
    is available for the latter three runs, which is shown in the inset in each
    of these three panels. The shaded area in that of the {\sc Bagpipes} run
    indicates the 16th to 84th percentile range.
}
    \label{fig:sed_fit}
\end{figure*}

    Table~\ref{tab:sed_fit} lists the most relevant physical parameters derived
from the SED analysis, and Figure~\ref{fig:sed_fit} shows the corresponding
template spectrum in each run superposed on the observed SED. For most of the
parameters, the four widely different SED fitting runs give largely consistent
results. The most robust properties obtained are the stellar mass and the age,
which are also the most critical in understanding the stellar population
properties. Dz5289 appears to have stellar mass of $10^{9.5-10.0} M_\odot$,
with an age of 330--510~Myr (i.e., formation redshift $z_f \approx 7.0$--8.5).
The inferred SFR ranges from 8.2 to 18.0~$M_\odot$~yr$^{-1}$, which is broadly 
consistent with the lower limit of 14.8--16.1~$M_\odot$~yr$^{-1}$ derived from 
the measurements of the H$\alpha$ line obtained from only part of the galaxy
(see Section 3). The derived dust extinction $A_{\rm V}$ for stars range from 
0.3 to 1.0~mag, or equivalently, ${\rm E(B-V)_{star}} = 0.07$--0.25~mag. 
Considering ${\rm E(B-V)_{star}=0.44\times E(B-V)_{gas}}$, this is in
reasonable agreement with the value of ${\rm E(B-V)_{gas}}=0.30\pm0.20$~mag
obtained from the Balmer decrement (see Section 3). Not surprisingly, the
metallicity parameter has a very wide range because it is very difficult (if 
not impossible) to derive metallicity based on broad-band photometry. 

\begin{table*}[hbt!]
    \centering
    \caption{Fitted physical properties from SED analysis. }
    \begin{tabular}{cccccc} \hline
        Tool & EAZY-py & CIGALE & Bagpipes & Prospector \\ \hline 
        $\log_{10}(M/M_\odot)$ & $9.48^{+0.04}_{-0.03}$ & $9.53\pm0.15$ & $9.65^{+0.13}_{-0.15}$ & $10.01^{+0.17}_{-0.23}$ \\

        Age~[Gyr] & ... & $0.51\pm0.31$ & $0.33^{+0.33}_{-0.16}$ & $0.35^{+0.36}_{-0.16}$ \\

        SFR & $8.15^{+3.25}_{-0.85}$ & $17.99\pm10.09$ & $12.28^{+5.86}_{-4.25}$ & $14.26$ \\

        $A_V$ & $0.32_{-0.07}^{+0.20}$ & ... & $0.59^{+0.18}_{-0.19}$ & $1.04^{+0.16}_{-0.13}$ \\

        $E(B-V)_{\rm star}$ & ... & $0.19\pm0.05$ & ... & ... \\

        $Z/Z_\odot$ & ... & $1.77\pm1.53$ & $1.45^{+0.75}_{-0.86}$ & $0.03^{+0.09}_{-0.02}$ \\
        \hline
    \end{tabular}
    \raggedright
    \tablecomments{For {\sc EAZY}-py, the best-fit parameters are listed. For other
    three tools, the 50th percentile posterior values are reported.
    Star formation rates (SFRs) are in units of 
    $M_\odot$~yr$^{-1}$. ${\rm E(B-V)_{star}}$ is the reddening for stars, and
    ${\rm E(B-V)_{star}=0.44\times E(B-V)_{gas}}$. For the Calzetti extinction
    law, ${\rm A_V = 4.05\times E(B-V)_{\rm star}}$. We adopt $Z_\odot=0.013$.
    Age and metallicity are not fitted parameters in our {\sc EAZY}-py run
    using the adopted templates.
    }
    \label{tab:sed_fit}
\end{table*}

\section{Discussion}

    While it remains to be verified whether Dz5289 is rotation-supported, 
it is certain that it has a regular disk morphology. This makes Dz5289 one of
the only two disk galaxies confirmed at $z>5$ to date, with its redshift
($z=5.289$) being only slightly lower than that of Twister-z5 ($z=5.3$). Dz5289
is slightly smaller: it has a stellar mass of
(0.3--1.0)$\times 10^{10} {\rm M_\odot}$ and its disk extends $\sim$6.2~kpc in 
diameter (with $R_e\approx 1.3$~kpc), while Twister-z5 has a stellar mass of 
$\sim$$2.5\times 10^{10} {\rm M_\odot}$ and its disk is $\sim$9~kpc in diameter
($R_e\approx 2.3$~kpc). The most important difference between them is in their 
morphologies: Twister-z5 has a very strong bulge ($\sim$3~kpc in diameter) and 
a highly asymmetric disk, while Dz5289 is a very regular disk with little
sign of a bulge. Such a difference can only be caused by their different 
evolutionary routes, which should be investigated by future theories of disk
galaxy formation.

    Currently, the most critical issue is whether such an early time of disk
galaxy emergence can be explained in the conventional $\Lambda$CDM framework.
\citet{Tamfal2022} carried out GigaEris, a very high-resolution cosmological 
zoom-in simulation of a Milky Way-sized halo from $z=300$ to $z=4.4$, and they 
detected a flattened stellar disk component at various times since $z\sim 7$--8.
The follow-up study of this simulation by \citet{vanDonkelaar2024} identified 11
distinct stellar thin disks formed at different times along different accretion 
planes, with the earliest one appearing at $z\approx 7.2$. Their criteria for a
disk are that the ratio between the rotation and random velocities is larger 
than unity and that the feature should last for at least 150~Myr. An earlier
disk would disappear because its constituent stars would migrate inward;
however, a disk would reappear at a later time when new stars are born. 
Interestingly, the stellar disks that they found in GigaEris all have high
stellar mass of $\sim$1$\times10^{10}M_\odot$.

    At the first glance, the GigaEris results seem to suggest that an early 
stellar disk such as Dz5289 could be explained without introducing any unusual
physical mechanisms in galaxy formation. However, Dz5289 has an important
property that is different from the GigaEris disks: it is a large disk galaxy
of $\sim$3.1~kpc in radius, while the GigaEris disks found untill the end of
the simulation ($z=4.4$) are all confined to $\lesssim$1~kpc in radius. In this
regard, it is still not clear whether early disk galaxies such as Dz5289 are
compatible with the current theories, and further investigations will be needed.

\section{Summary}

    While fast-rotating gas disks are known to exist at high redshifts (up to 
$z=7.31$; see \citealt{Rowland2024}), it is unclear whether there are stellar
disks at comparable redshifts. In fact, most of the early gas disks reported in
the literature either do not show disk stellar counterparts or do not yet have
their stellar counterparts revealed. On the other hand, the JWST observations
over the past two years indicate that there could be abundant stellar disks at
$z>3$. The vast majority of these studies, however, rely on photometric
redshifts, and the confirmed cases are still scarce. 

   This work shows that D-CEERS-RUBIES-z5289 (Dz5289) is an edge-on disk galaxy
at $z=5.289$, adding the second confirmed $z>5$ stellar disk after Twister-z5
(at $z=5.3$). The current spectroscopic data do not allow the derivation of 
its rotation speed because the slit only covered part of the galaxy on one
side.  Nevertheless, the large width of the H$\alpha$ line 
($\sim$345~km~s$^{-1}$) indicates that the rotation speed is likely significant.
Its disk morphology is quantitatively verified in four NIRCam bands that 
correspond to the rest frame wavelength range of 0.44--0.70~$\mu$m, which shows
that the disk extends $\sim$6.2~kpc with $R_e$ of $\sim$1.3--1.4~kpc. 
Such a large stellar disk only $\sim$1.1~Gyr after the Big Bang is yet to be 
seen in numerical simulations and likely requires new models. To put more 
constraints on theories, a proper analysis of its kinematics would be
desirable. This could be enabled by new observations through the JWST integral
field spectroscopy.
\\

All the JWST data used in this paper can be found in MAST: 
\dataset[10.17909/bkqy-aj40]{http://dx.doi.org/10.17909/bkqy-aj40}.

\begin{acknowledgements}

We thank the referee for the constructive comments. We also thank Dr. de Graaff
for the helpful discussions. This work is based on the observations made with
the NASA/ESA/CSA James Webb Space Telescope and obtained from the Mikulski 
Archive for Space Telescopes, which is a collaboration between the Space 
Telescope Science Institute (STScI/NASA), the Space Telescope European
Coordinating Facility (ST-ECF/ESA), and the Canadian Astronomy Data Centre
(CADC/NRC/CSA). 

H.Y. and B.S. acknowledge the support from the University of Missouri Research 
Council grant URC-23-029 and the NSF grant AST-2307447. 
C.L. acknowledges the support from the Special Research Assistant program of 
the Chinese Academy of Sciences (CAS). 

\end{acknowledgements}

\bibliographystyle{aasjournal}

\end{document}